\documentclass[iop,apj]{emulateapj}

\usepackage{graphicx}         

\newcommand \kms{km~$\rm{s}^{-1}$}
\newcommand \cc{$\rm{cm}^{-3}$}
\newcommand \mum{$\mu$m}

\slugcomment{Submitted 2013 November 27; accepted  2014 July 21}

\begin{document}

\shorttitle{Proper Motions in the Mid-IR}
\shortauthors{Noriega-Crespo et al.}


\def \kms{km~$\rm{s}^{-1}$}
\def \cc{$\rm{cm}^{-3}$}
\def \mum{$\mu$m}

\title{Proper Motions of Young Stellar Outflows in the Mid-Infrared with
Spitzer.\hskip 40pt {II. HH 377/Cep E}}

\author{Noriega-Crespo, A.\altaffilmark{1,3},
Raga, A. C. \altaffilmark{2},
Moro-Mart\' \i n, A.\altaffilmark{3},
Flagey, N.\altaffilmark{4,6},
and
Carey, S. J.\altaffilmark{5}}

\altaffiltext{1}{Infrared Processing and Analysis Center,
California Institute of Technology, CA, 91125, USA}
\altaffiltext{2}{Instituto de Ciencias Nucleares, Universidad Nacional 
Aut\'onoma de M\'exico, Ap. 70-543, 04510 D.F., M\'exico}
\altaffiltext{3}{Space Telescope Science Institute, 3700 San Martin Dr.,
  Baltimore, MD, 21218, USA}
\altaffiltext{4}{Jet Propulsion Laboratory, California Institute of 
Technology, CA, 91099,  USA}
\altaffiltext{5}{Spitzer Science Center, California Institute of 
Technology, CA, 91125,  USA}
\altaffiltext{6}{Institute for Astronomy, 640 North A\'ohoku Place,
  Hilo, HI, 96720, USA}

\begin{abstract}

\smallskip

We have used multiple mid-infrared observations
at 4.5~\mum~obtained with the Infrared Array Camera,
of the compact ($\sim$ 1.\arcmin4) young stellar bipolar 
outflow Cep E to measure the proper motion of its brightest
condensations. The images span a period of $\sim 6$ yr and
have been reprocessed to achieve a higher angular resolution 
($\sim$ 0.\arcsec8) than their normal beam ($\sim$ 2\arcsec).

We found that for a distance of 730 pc, the tangential velocities
of the North and South outflow lobes are 62$\pm$29 
and 94$\pm$26~\kms~respectively, and moving away from the central
source roughly along the major axis of the flow.  
A simple 3D hydrodynamical simulation of the
H$_2$ gas in a precessing outflow supports this idea. Observations and
model confirm that the molecular Hydrogen gas, traced by the 
pure rotational transitions, moves at highly supersonic velocities  
without being dissociated. This suggests either a very
efficient mechanism to reform H$_2$ molecules along these shocks
or the presence of some other mechanism (e.g. strong magnetic field)
that shields the H$_2$ gas.

\end{abstract}

\section{Introduction}

The measurement of proper motions for Herbig-Haro outflows and stellar
jets has a long tradition and has played a fundamental role in our
understanding of early phases of evolution of low mass stars,
including their accretion rates, mass loss and disk dissipation 
(see e.g. Mckee \& Ostriker 2007; Bally 2009). 
The original work by Herbig \& Jones (1981) on the first and brightest
Herbig-Haro (HH) objects 1, 2 and 3, using photographic plates
over a 34 yr period, set up the framework of the mass loss process in proto-stars
 and their interaction with the surrounding medium. The ``knots'' of
 HH 1 and 2 were found to have tangential velocities ranging from 100
 to 350~\kms, in disagreement with the spectroscopic measurements that
 indicated that their emission was due to shocks of at most 
100~\kms~(see e.g. Raymond 1979).
Modern observations and models have shown that HH 1 and 2 are
 the leading working surfaces or 'bowshocks' of highly collimated
 jet/counter-jet system that arises from a deeply embedded
 protostellar source of Class 0. And that a time dependent ejection
 can account for both larger proper motion and relative smaller shock 
velocities between their knots (see e.g. Raga et al. 2011a for a review).
At optical wavelengths, using narrow band images at some of the
brightest  collisionally excited emission lines (e.g. H$\alpha$ and/or
[SII]), it has been possible to measure proper motions for $\sim 50$
Herbig-Haro flows, within a distance of  roughly one kiloparsec,
 both from the ground and space (see e.g. Bally, Reipurth \& Davis 2007; 
Caratti o Garatti \& Eisl\"offel 2009) and over relative short periods of time 
($\le$ 10--20 yr). In the case of objects like HH 1/2, 34 and 46/47, 
such measurements have led to some spectacular time sequences of the
outflows  (see e.g.  Hartigan et al. 2011). Proper motions are still
a fundamental tool to find and correlate outflows across the sky and 
over distance scales of parsecs (see e.g. Reipurth et al. 2013). This method
is particularly important when the outflow driving source has not
been clearly identified. Radio observations have also been very successful in 
measuring proper motions of outflows for low and high 
mass protostars, taking advantage of the gas flow thermal (free-free)
emission and high angular resolution measurements obtained by interferometric
observations (see e.g. Rodriguez 2011).

At infrared wavelengths, however, it has been more difficult to 
measure proper motions because the lack of large format arrays.
 It has been until recently that near/mid-IR arrays have had a wide
 enough FOV to include a reasonable number of reference stars and 
be able to cross-correlate multiple epoch observations to derive the 
proper motions. The large proper motions observed in HH 1/2 system in
the atomic ionic gas (Herbig \& Jones 1981) have been measured in the
NIR as well, using the molecular Hydrogen shock excited emission 
vibrational transition (v=1-0 S(1)) at 2.121~\mum~(Noriega-Crespo et
al. 1997). It is now possible, using similar tracers, to measure 
the motions of multiple H$_2$ features in Cha II (Caratti o
Garatti et al. 2009)   or  $\rho$ Oph clouds (Zhang et al. 2013),
with a similar range of transversal velocities (from 30 to 120 \kms).

In the mid-IR, thanks to the stability and longevity
of the Infrared Array Camera (IRAC; Fazio et al. 2004) on board the
Spitzer Space Telescope, it has been possible to measure the proper
motions of several outflows in NGC1333 at an angular 
resolution of 2\arcsec~(Raga et al. 2013; Paper I).  IRAC channel 2 at
4.5~\mum~is an excellent tracer of the H$_2$ rotational emission, 
since three of the brightest lines, 0-0 S(11) 4.18, 0-0 S(10) 4.40  and
0-0 S(9) 4.18~\mum~fall within its passband (Noriega-Crespo et al. 2004a,b; 
Looney et al. 2007; Tobin et al. 2007; Ybarra \& Lada 2009; 
Raga et al. 2011a, 2012; Noriega-Crespo \& Raga 2012).

Using images obtained at 4.5~\mum~over a period of $\sim$ 7 yr, 
Raga et al. (2013) obtained tangential velocities ranging from 
$\sim 10$ to 100~\kms~(for a 220 pc distance) 
for 8 outflows in NGC~1333 cloud.
For the bright HH 7-11 system, that lies at the center of the cloud,
the H$_2$ tangential velocities of $\sim 10$--15 km s$^{-1}$
are very close to those obtained at optical wavelengths using atomic gas 
line tracers, such as [SII] 6717/31\AA~ or H$\alpha$ (Herbig \& Jones 1983; 
Noriega-Crespo \& Garnavich 2001).

In summary, proper motions, and the corresponding tangential
velocities, are essential for determining the dynamics of the
outflows, plus
their momentum and energy transfer into the surrounding interstellar
medium (see e.g. Quillen et al. 2005; Padoan et al. 2009;
Plunkett et al. 2013). The striking morphological
similarity between the atomic/ionic gas emission (obtained from optical
or near-IR observations) and that of the molecular Hydrogen (obtained
either from near/mid-IR observations), suggests that the
kinematics of the protostellar outflows
allow this relatively fragile molecule ($H_2$) either to survive the
shocks or to regenerate itself rapidly in the dense postshock regions
(see e.g. Le Bourlot et al. 2002; Panoglou et al. 2012).
These issues can be partially addressed with studies of proper motions
of the H$_2$ emission from stellar outflows.

In this study we determine the proper motions of the deeply 
embedded and compact molecular outflow Cep E, driven by a
intermediate mass class 0 protostar
(Eisl\"offel et al. 1996;  Lefloch, Eisl\"offel
\& Lazareff 1996; Ladd \& Hodapp 1997,
Noriega-Crespo, Garnavich \& Molinari 1998; 
Hatchel, Fuller \& Ladd 1999; Moro-Mart\'\i n et al 2001; 
Noriega-Crespo et al. 2004b). Cep E is considered
an excellent prototype of its kind, and therefore, has
prompted many recent observations at millimeter and 
sub-millimeter wavelengths to study 
its H$_2$O and CO molecular content (Lefloch et al. 2011; G\'omez-Ruiz
et al. 2012; Tafalla et al. 2013), in part because its 
shocked excited South lobe is detected at optical 
wavelengths (HH~377; (Ayala et al. 2000). 

At a distance of $\sim$ 730 pc to Cep E, measuring proper motions with
a time interval of $\sim 6$ yr (covered by the
available IRAC images) is a considerable
challenge, since velocities of $\sim 100$~\kms~would correspond to
shifts of only $\sim 0.17$\arcsec. In order to achieve as high a resolution
as possible, we have employed a high
angular resolution enhancement of the IRAC images, 
reaching a resolution of 0.6\arcsec--0.8\arcsec~(see Velusamy, Langer
\& Marsh 2007; Velusamy et al. 2008; Noriega-Crespo \& Raga 2012; 
Velusamy, Langer \& Thomson 2014).
Such an enhancement has recently been successfully applied
to IRAC images of Cep E  (Velusamy et al. 2011).  Finally, given that
observationally there is a tremendous morphological similarity between 
the mid-IR and NIR emission (Noriega-Crespo et al. 2004b),
we expand the study to include some ground based H$_2$ 2.12~\mum~NIR data 
that allows us to extend the time baseline of the observations
to $\sim 16$ yr (Table 1).

The paper is organized as follows. The observations and their
high angular resolution  reprocessing is described in $\S$ 2.
The determination of the proper motions is described in $\S$ 3.
Finally, $\S$ 4 presents a summary of the results.

\section{Observations \& High Angular Resolution Reprocessing}

The Cep E outflow was one of the first young stellar outflows observed 
with the Spitzer Space Telescope as part of their Early Release Observations 
(ERO; PID 1063, P.I.  Noriega-Crespo) because of its strong brightness 
at mid-IR wavelengths. Its emission in the mid-IR (5--17~\mum) is due mostly 
to bright H$_2$ rotational lines clearly detected already by the Infrared 
Space Observatory (ISO) with the infrared camera (ISOCAM) using its Circular 
Variable Mode (CVF; see e.g. Boulanger et al. 2005) 
that provided a low spectral (R$\sim$ 45) and angular (FWHM$\sim$ 6\arcsec)  
resolution spectral map of the region (Moro-Mart\' \i n et al. 2001). 
The outflow was later observed by two ambitious programs, one to map the 
Cepheus OB3 molecular cloud to study its star formation 
(PID 20403, P.I. Pipher), and more recently by the GLIPMSE360 
survey during the Warm  Spitzer phase as one of the  large Exploration 
programs (PID 60020, P.I. Whitney). 
The data, consisting of the basic calibrated frames or BCDs,
have been recovered from the Spitzer Legacy Archive, version S18.18.0 (Cryo) 
and S19.0.0 (Warm). In all cases, the data was collected using the
High-Dynamic-Range (HDR) mode with a 12 s integration time for the 'long' 
frames (10.4 s on target) and 0.6 s for the 'short' ones. A summary of the observations 
is presented in Table 1. 

\begin{deluxetable}{rcccccc}
\tablecaption{Cep E Observations\label{tbl:colors}}
\tablewidth{0pt}
\tablecolumns{6}
\tablehead{
\colhead{Program} & \colhead{Request}  & \colhead{Observing} 
& \colhead{Mean} & \colhead{t$_{exp}$} &\colhead{Observer}\\
\colhead{ID} & \colhead{Key}  & \colhead{Date} 
& \colhead{Coverage} & \colhead{(sec)} & \colhead{}}
\startdata
\nodata & \nodata  & 1996-08-22 & \nodata & 1800. & Noriega-Crespo\tablenotemark{a}\\
 1063   & 6064384  & 2003-11-26 & 9 & 93.6 & Noriega-Crespo\tablenotemark{b}\\
 20403  & 15571968 & 2005-12-25 & 4 & 41.6 & Pipher\tablenotemark{b}\\
 20403  & 15570432 & 2006-08-09 & 4 & 41.6 & Pipher\tablenotemark{b}\\
 60020  & 38734592 & 2010-02-01 & 8 &  83.2 & Whitney\tablenotemark{b,c}\\
\nodata & \nodata  & 2012-08-29 & \nodata & 1500. &Flagey+\\
\nodata & \nodata  & \nodata & \nodata & \nodata & Noriega-Crespo\tablenotemark{a}\\
\enddata
{\baselineskip=0pt
\tablenotetext{a}{Ground based 2.12~\mum}
\tablenotetext{b}{IRAC 4.5~\mum}
\tablenotetext{c}{Warm Spitzer Observation}}
\end{deluxetable}

The BCDs were then reprocessed with the HiREs deconvolution software 
AWAIC\footnote[3]{http://wise2.ipac.caltech.edu/staff/fmasci/awaicpub.html}
(A WISE Astronomical Image Co-Adder), developed by the Wide Field Infrared  
Survey Explorer (WISE) for the creation of their Atlas images (see e.g.
Masci \& Fowler 2009; Jarrett et al. 2012). 
The AWAIC software optimizes the coaddition of individual frames by making
use of the Point Response Function (PRF) as an interpolation kernel, to avoid
flux losses in undersampled arrays like those of IRAC, and also
allows a resolution enhancement (HiRes) of the final image, by removing its
effect from the data in the deconvolution process. We have used this
method quite successfully in the HH 1/2 outflow 
(Noriega-Crespo \& Raga 2012), and as mentioned above, 
a similar method has been used on Cep E 
(Velusamy et al. 2011) and  HH 46/47 (Velusamy et al. 2007).  
On IRAC images, the HiRes  enhances the angular resolution from the 
standard $\sim 2$\arcsec~to $\sim$ 0.6\arcsec--0.8\arcsec~
(Velusamy et al. 2008; Noriega-Crespo \& Raga 2012).

The combination of being deeply embedded and its youth, $\sim 5000$ yr 
(Ladd \& Hodapp 1997), perhaps makes Cep E one of the outflows where the 
morphology of the vibrational H$_2$ is nearly identical to the H$_2$ 
rotational emission observed with IRAC at 4.5~\mum. This similarity 
has encouraged us to introduce an earlier H$_2$ v=1-0 2.12~\mum~image 
from 1996 and one recently obtained in 2012, in the analysis 
of the proper motions, providing a $\sim$ 16 yr time baseline. 
The NIR 1996 image was obtained with the 3.5 m telescope at the 
Apache Point Observatory with a 256$\times$256 array at f/5 
and a 0.\arcsec482 pixel$^{-1}$ scale using a 2.12~\mum~filter 
(1\% width) and 2.22~\mum~(4\% width) to subtract the continuum.
The complete analysis of these data was already presented by 
Ayala et al. (2000). The 2012 image was obtained at Palomar 
Observatory with the Wide Field Infrared Camera (WIRC) mounted in the 
200in prime focus using a 2048 x 2048 Hawaii-II HgCdTe detector. WIRC has
a field-of-view of 8.\arcmin7 and a 0.2487 arcsec/pixel scale 
(Wilson et al. 2003). The observations were carried out on August 29, 
2012 in the 2.12~\mum~and K-continuum 
(2.27~\mum, 2\%) filters, with a 25 min total 
integration time. Figure 1 shows a comparison of the Palomar 
2.12~\mum~continuum subtracted image with that at 4.5~\mum~from 
Noriega-Crespo et al. (2004b). Among some of the small obvious 
differences are the lack of H$_2$ vibrational emission on the same 
region where there is a ``wide angle'' cone at 4.5~\mum~
(Velusamy et al. 2011), that strongly suggest to be scattered 
light by small dust particles; and the emission at 2.12~\mum~on the South 
lobe that fits within these ``cones'' and reaches further into the 
IRAS 23011+6126 central source. Other than these differences, the knots 
that we have used in our proper motion 'boxes' are indistinguishable 
from each other. In Figure 2 we show the five epochs that are being 
analyzed; the IRAC images are the HiRES AWAIC reprocessed 
after 60 iterations.

\begin{figure}
\centering
\includegraphics[width=1\columnwidth,angle=0]{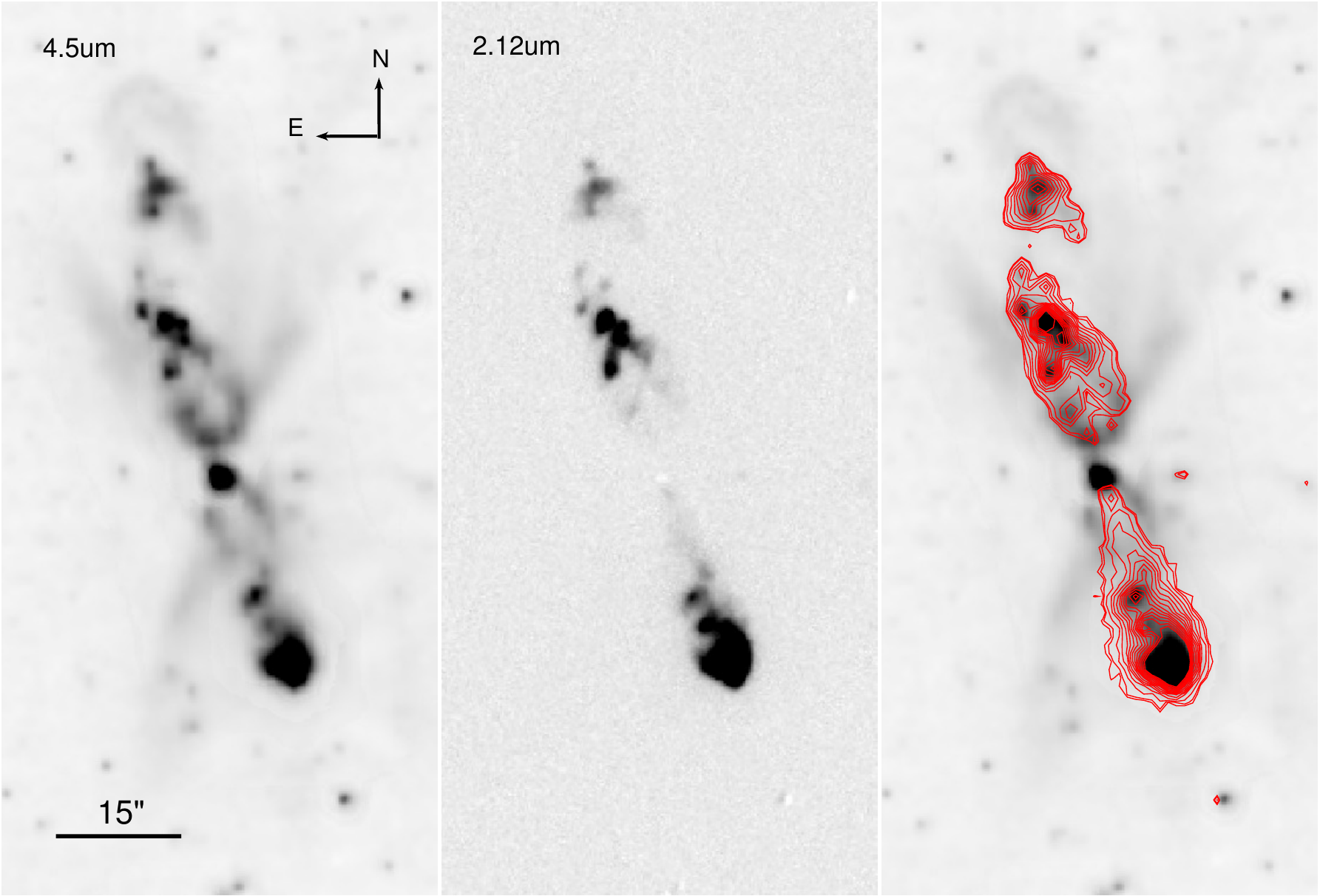}
\caption{A comparison of the rotational H$_2$ emission as detected
by IRAC at 4.5~\mum~in 2003.89 (left and right panels) 
with that from ground based observation of the vibrational 
H$_2$ v = 1-0 at 2.12~\mum~(central panel and red contours
on the IRAC 4.5~\mum~image) obtained in 2012.66. From this comparison
it looks that the 4.5~\mum~image includes in its cavity a significant component
of continuum emission due to dust scattered light.\label{fig1}}
\end{figure}

\begin{figure*}
\epsscale{1.0}
\centering
\includegraphics[width=1.7\columnwidth,angle=0]{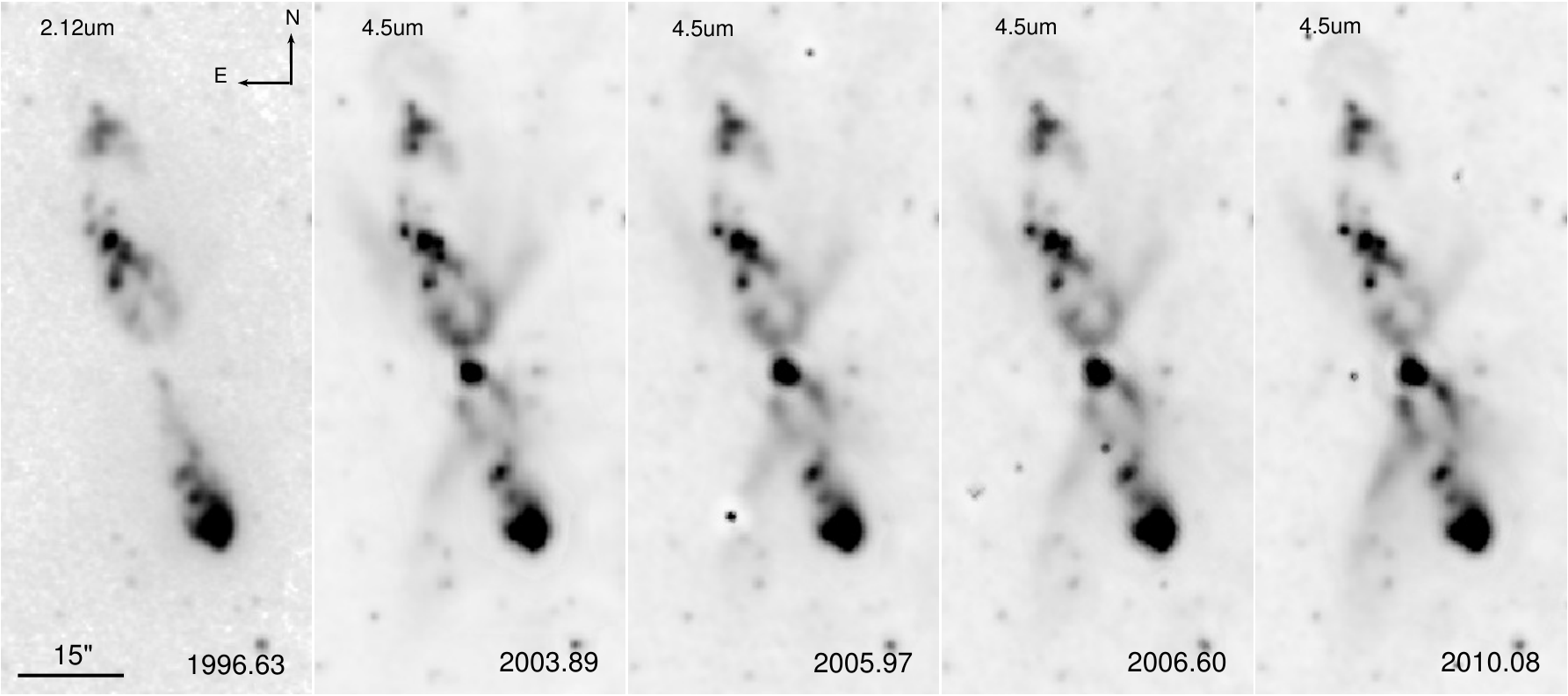}
\caption{The other 5 epochs of Cep E data used to determine the 
NIR/midIR proper
motion of its knots, including the 1996.63 2.12~\mum~ground based image 
(first frame from left to right), plus the
other 4 epochs of the AWAIC HiRES IRAC images at 4.5~\mum, 
after 60 iterations (see Table 1).\label{fig2}}
\end{figure*}

\section{Proper Motion Measurements}

As can be seen in Figure 2, the Cep E outflow shows bipolar
``cavity'' structures which extend out to $\sim 10$--15\arcsec~
from the source (with an approximately NNE-SSW orientation).
These cavities have two ridges that first open out from the
outflow source, and then converge into compact emission structures.
Further away ($\sim 20$\arcsec~from the source), we find two
bow-like prolongations of the outflow lobes.
As can be seen in Figure 2, the emission along the
two-ridged cavities shows a complex time-dependence, with the
Northern cavity becoming fainter and the Southern cavity
brightening from 2003 to 2010 (i.e., the period covered
by the IRAC images).

We have defined four boxes, including the regions of convergence 
of the cavities and the bow-like structures, which we show in Figure 4.
For each of these four boxes, we have carried out cross correlations
between the emission in the 2003.89 frame (the first of the IRAC
frames, see Table 1) and the other 4 available frames. From paraboloidal
fits to the peak of cross correlation functions we then determine the offsets
of the emission (within the four boxes, see Figure 4) with respect to
the 2003.89 frame. For the two 2.12~\mum~images, we performed the
cross correlation initially with 2003.89 frame as well, 
and then between themselves. In this way the offsets at 2.12 \& 4.5~\mum~are 
measured on a common reference frame.


\begin{deluxetable}{crrrrr}
\tablecaption{Cep E Proper Motions and Tangential Velocities
from 2.12~\mum~images\tablenotemark{a}\label{tbl:colors}}
\tablewidth{0pt}
\tablecolumns{4}
\tablehead{
\colhead{Box} & \colhead{$\Delta_x$}\tablenotemark{b}  &
\colhead{$\Delta_y$}\tablenotemark{b} & \colhead{V$_x$}\tablenotemark{b}
& \colhead{V$_y$}\tablenotemark{b} & \colhead{V$_T$}\\
\colhead{} & \colhead{mas/yr}  & \colhead{mas/yr} & \colhead{\kms}
& \colhead{\kms} & \colhead{\kms}}
\startdata
1  &  4.0  &  -8.7 &  13.9 & -30.2 & 33.0$\pm$14.6\\
2  &  0.9  &  -2.6 &   3.1 &  -9.0 &  9.5$\pm$14.0\\
3  & -0.6  &   6.7 &  -2.1 &  23.3 & 23.4$\pm$11.9\\
4  &  0.2  &  12.2 &   0.7 &  42.4 & 42.4$\pm$11.2\\
\enddata
{\baselineskip=0pt
\tablenotetext{a}{For a distance of 730 pc}
\tablenotetext{b}{The proper motions have estimated
errors of 3.2 mas/yr (11 km s$^{-1}$)}}

\end{deluxetable}

\begin{deluxetable}{crrrrr}
\tablecaption{Cep E Proper Motions and Tangential Velocities
from IRAC 4.5~\mum~images\tablenotemark{a}\label{tbl:colors}}
\tablewidth{0pt}
\tablecolumns{4}
\tablehead{
\colhead{Box} & \colhead{$\Delta_x$}  & \colhead{$\Delta_y$} 
& \colhead{V$_x$} & \colhead{V$_y$} & \colhead{V$_T$}\\
\colhead{} & \colhead{mas/yr}  & \colhead{mas/yr} & \colhead{\kms}
& \colhead{\kms} & \colhead{\kms}}
\startdata
1  &  14.7$\pm$9.3  & -10.4$\pm$1.6 & 51.1$\pm$32.3 & -36.2$\pm$ 5.6 & 62.6$\pm$29.5\\
2  &  10.0$\pm$1.4  &  -7.3$\pm$6.8 & 34.8$\pm$ 4.9 & -25.4$\pm$23.6 & 43.1$\pm$17.9\\
3  &   2.7$\pm$2.9  &  20.4$\pm$8.1 &  9.4$\pm$10.1 &  70.9$\pm$28.2 & 71.5$\pm$29.3\\
4  & -12.8$\pm$6.0  &  23.8$\pm$5.4 &-44.5$\pm$20.9 &  82.8$\pm$18.8 & 94.0$\pm$26.5\\
\enddata
{\baselineskip=0pt
\tablenotetext{a}{For a distance of 730 pc}}
\end{deluxetable}

\begin{figure}
\epsscale{0.8}
\plotone{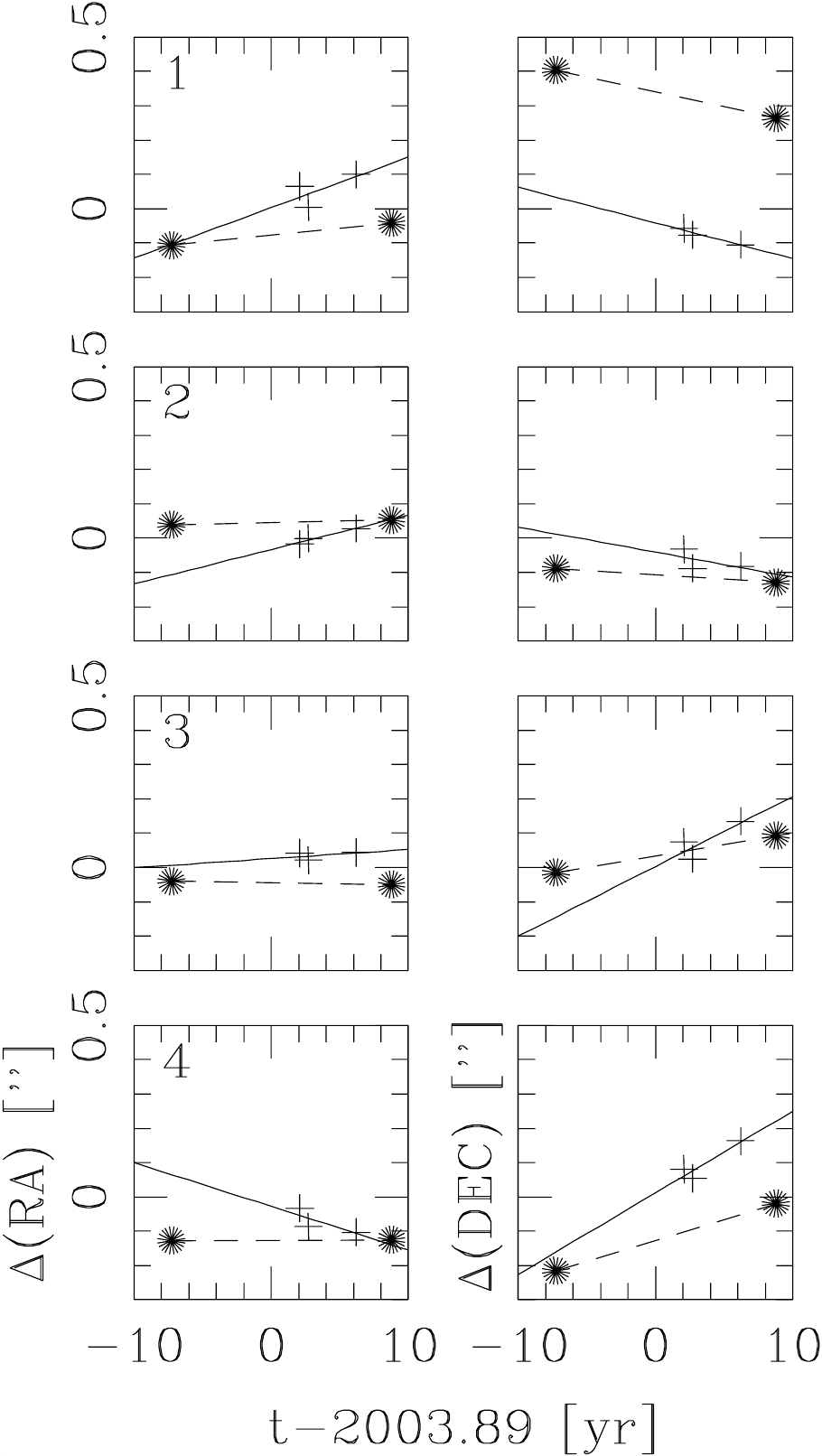}
\caption{The offsets in arcseconds in RA and DEC between the 
5 epochs of Cep E as measured in 4 different boxes; two each 
for the North and South lobes of the outflow (see Figure 4). 
The symbols correspond to the 2.12\mum~(stars) and 4.5\mum~(plus signs)
data, respectively.}
\label{fig3}
\end{figure}

The resulting RA and DEC displacements for the four selected
boxes are plotted as a function of time in Figure 3, where the stars 
and plus signs correspond to the 2.12 \& 4.5~\mum~frames
respectively. From this Figure, one
notices that boxes 1 and 4 (corresponding to the bow-like
structures, see Figure 4) show substantial N-S motions,
while the ``cavity tips'' (boxes 2 and 3) have considerably
lower proper motions. We have then carried out linear fits
to the time dependencies of the RA and DEC offsets of
our four boxes, from which we obtain the proper motions
(and their associated errors) given in Tables 2 (2.12~\mum) 
\& 3 (4.5~\mum).

\begin{figure}
\epsscale{1.0}
\plotone{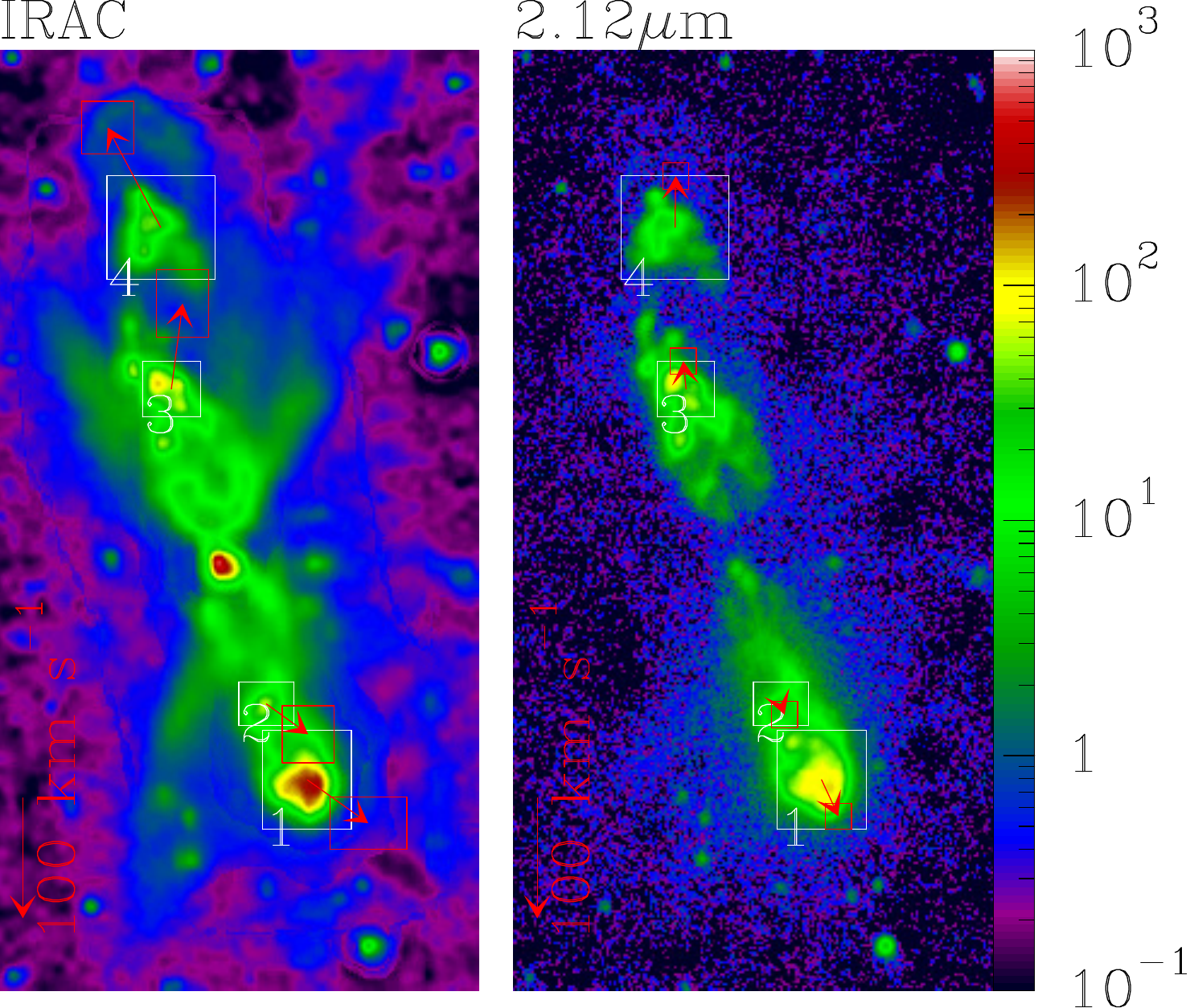}
\caption{ The tangential velocities of the H$_2$ 
emission as detected in IRAC 4.5~\mum~(left) and 2.12~\mum~(right)
in the four boxes used in this study (see Table 2 \& 3).}
\label{fig4}
\end{figure}

The proper motions obtained from the 2.12~\mum~and IRAC 4.5~\mum~do 
show similar trends and are within a factor two (Table 2).
One interesting thing to notice is that by
placing all displacements on the same frame of reference, 
one can measure also a shift between the positions of the 
vibrational (2.12~\mum) and rotational (4.5~\mum) H$_2$ emission
in the outflow. These are of the order of 0.25\arcsec--0.30\arcsec, 
certainly larger than the 0.05\arcsec--0.10\arcsec positional uncertainty
of the selected regions. 
The H$_2$ proper motions show evidence, for first time, that
both the vibrational and rotational H$_2$ can share the same kinematics 
within an outflow; and furthermore, that the vibrational and
rotational H$_2$ gas do trace slightly different regions within the
outflow. Although this last statement may seem obvious within the now classical
scenario of the acceleration of a molecular outflow, where the atomic/ionic 
supersonic flow drags and/or excites the surrounding molecular gas
(see e.g. Raga et al. 1995), this is (to our knowledge) 
one of the first observational evidence that this is the case 
for the different components of the H$_2$ gas. Recall that 
spectroscopically we do have some good examples where both near-IR
and mid-IR H$_2$ emission lines have been observed simultaneously in 
an outflows. Perhaps one of best examples is that of OMC-1 
(Rosenthal, Bertoldi \& Drapatz 2000) where the Short Wavelength 
Spectrometer (SWS) detected 56 H$_2$ transitions within its 
2.5 to 45~\mum~spectral range, and where the rotational emission was 
tracing a H$_2$ gas with an excitation temperature of 600 K, while 
the vibrational emission was tracing one at 3200 K. In OMC-1 
it was possible to explain the bulk of the emission with collisional 
excitation produced by a combination of C-type and J-type
shocks (Rosenthal, Bertoldi \& Drapatz 2000). In Cep E, 
the spectroscopic evidence (Moro-Martin et al. 2001) 
also suggest a mixture of excitations to explain the optical 
and near/mid/far IR observations.

The resulting tangential velocities are shown (together with
their error boxes) in Figure 4. Because of the multiple epochs, the
motions seem to be better defined at 4.5~\mum~than at 2.12~\mum.
At 4.5~\mum, the bow-like structures
show proper motions of 94~\kms~(Southern bow, box 1) and
63~\kms~(Northern bow, box 4) directed approximately away from
the outflow source. The Southern cavity tip (box 2) has essentially
zero proper motion in the X-axis, and if this is the case it
follows the flow of the bowshock but at steeper angle. 
The Northern cavity tip (box 3) shows a proper motion 
of 44~\kms~at an angle of
$\sim 60^\circ$ from the outflow direction (see Figure 4).
This proper motion measurement might be affected by 
significant intensity variations of the cavity emission during
the observed time period (see Figure 2).

\section{Numerical Simulations}

In one of the first studies of the H$_2$ NIR emission in Cep E 
(Eisl\"offel et al. 1996) the authors suggested that the ``wiggles and 
sideways positional offsets'' were due to precession, with a relatively 
small precession angle of 4\arcdeg. In the same study they noticed the 
presence of a couple of H$_2$ knots emanating westward from the central 
source and nearly perpendicular to the main Cep E outflow 
(Eisl\"offel et al. 1996, Figure 3), suggesting a very close-by
second protostar, and therefore, a possible mechanism to drive 
the precession. To further test this hypothesis and compare with the 
kinematical behavior of the H$_2$ gas derived from the proper motion 
measurements, we present in this section some relatively simple 3D 
hydrodynamical simulations. The YGUAZU-A code (Raga et al. 2000, 
Raga et al. 2003) was selected for this simulation.
The code, in a nutshell, uses a binary adaptive grid and integrates the
gas-dynamic equations with a second-order accurate scheme (in time and space)
using a flux-vector splitting method (van Leer 1982). The code has been 
used over a decade to simulate the gas dynamical processes that take place 
in several astrophysical scenarios including YSO outflows (e.g. Raga et al. 
2004), proto-planetary nebulae (Vel\'azquez et al. 2011), 
supernova explosions (Vel\'azquez et al. 2004),
photodissociation regions (Reyes-Iturbide et al. 2009), 
photoevaporating clumps (Raga, Steffen \& Gonzalez 2005), 
and MIRA's turbulent wake (Raga et al. 2008) among other. 
In this version of the code we include a H$_2$ gas component, 
although we cannot distinguish between its vibrational/rotational excitation.

\begin{figure*}
\centering
\epsscale{1}
\includegraphics[width=1.7\columnwidth,angle=0]{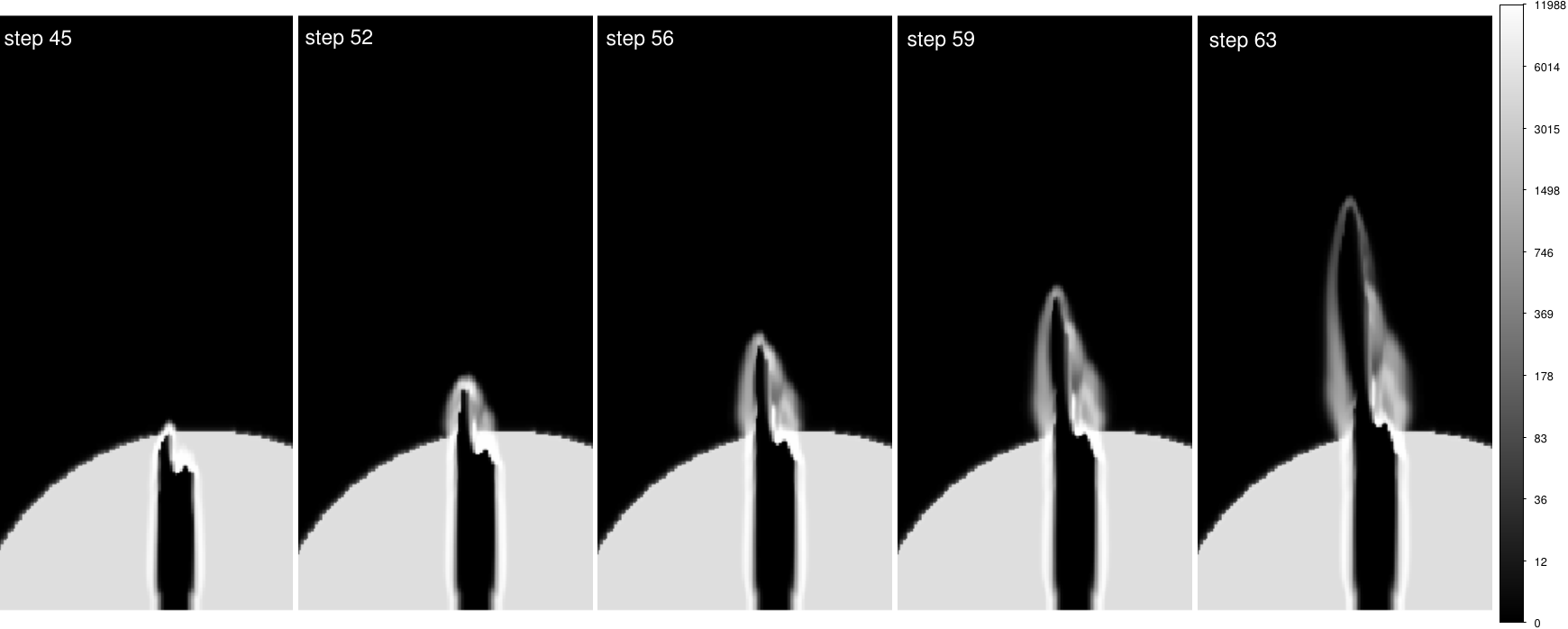}
\caption{The H$_2$ number density  [from 0 (black) 
to 1.2$\times 10^3$\cc~(white)] for a precessing jet emerging
from a cold core to simulate the Cep E outflow.
\label{fig5}}
\end{figure*}

The goal of the simulation is not to model in detail Cep E, but to show
that the observed proper motions can indeed be explained when taking
into account precession. Because of the relatively small dynamical age 
of the Cep E outflow ($\sim$ 3000 yr), we set the initial conditions of 
the model as that of a jet emerging from a compact dense cold core
with the simplifying assumption that it is in thermal
balance with the surrounding medium ($n_{core} = 1000$~\cc, $T_{core} = 1$ K,
$n_{ISM} = 10$~\cc, $T_{ISM} = 10^3$ K). Based on our previous simulations 
(Raga et al. 2004), we assume that the jet has an initial radius of 
$1.5\times 10^{16}$ cm, a temperature of 1000 K
and a velocity of 200~\kms, with a time dependent
velocity variation (``pulses'') of a 50\% over a 60 yr period. 
The jet is set to precess on a 10\arcdeg angle 
with respect to its cylindrical symmetry axis over a 1200 yr period.
Since we are interested on the bulk motion of the gas to compare with what
is observed in the proper motions, we  have chosen a medium resolution
computational grid of $128 \times 128 \times 256$ cells set 
to a scale of (X,Y,Z) = (5,5,10)$\times 10^{17}$ cm, respectively.
The time resolution is set to 200 yr per step, so once the jet 
plunges through the core, it takes 15 to 20 frames to reach a 
dynamical age close to that of Cep E.

The results of the simulation for the H$_2$ number density
 are shown for five time steps starting at the moment when 
the jet finally breaks free from the dense core
(Figure 5; step 45). In all cases we present the XZ projection, 
i.e. perpendicular and along the flow.
In the H$_2$ number density one already can see 
that the initial effect of precession has been to widen the path of 
the flow and the creation of two different density maxima upstream.
At about 2000--3000 yr later (steps 56 and 59),
the maxima have nearly merged and the flow is compact and asymmetric.

In an effort to better compare the numerical model with the observations
we have integrated in the 3D grid the H$_2$ emission along the line of sight
and projected it on a 30\arcdeg~angle. We have taken the difference 
in the projected emission for models 56 and 59 (i.e. a 600 yr interval), 
to mimic as much as possible the proper motion measurements. 
The result (Figure 6)  shows at least four ``knots'' with 
tangential velocities 
ranging from 10 to 50~\kms~on slightly different directions, although 
the bulk of the motion is away from the center of the grid. 
A 50~\kms~tangential velocity is certainly consistent with the value of 
$\sim$ 62~\kms~of the North lobe, that plunges deeper into the cloud,
and is a bit smaller than the 94~\kms~value of the South lobe, where 
one detects at optical wavelengths HH 377, i.e the observed proper
motions seem to reflect the difference in physical environment 
between the two lobes.

\begin{figure}
\epsscale{0.7}
\plotone{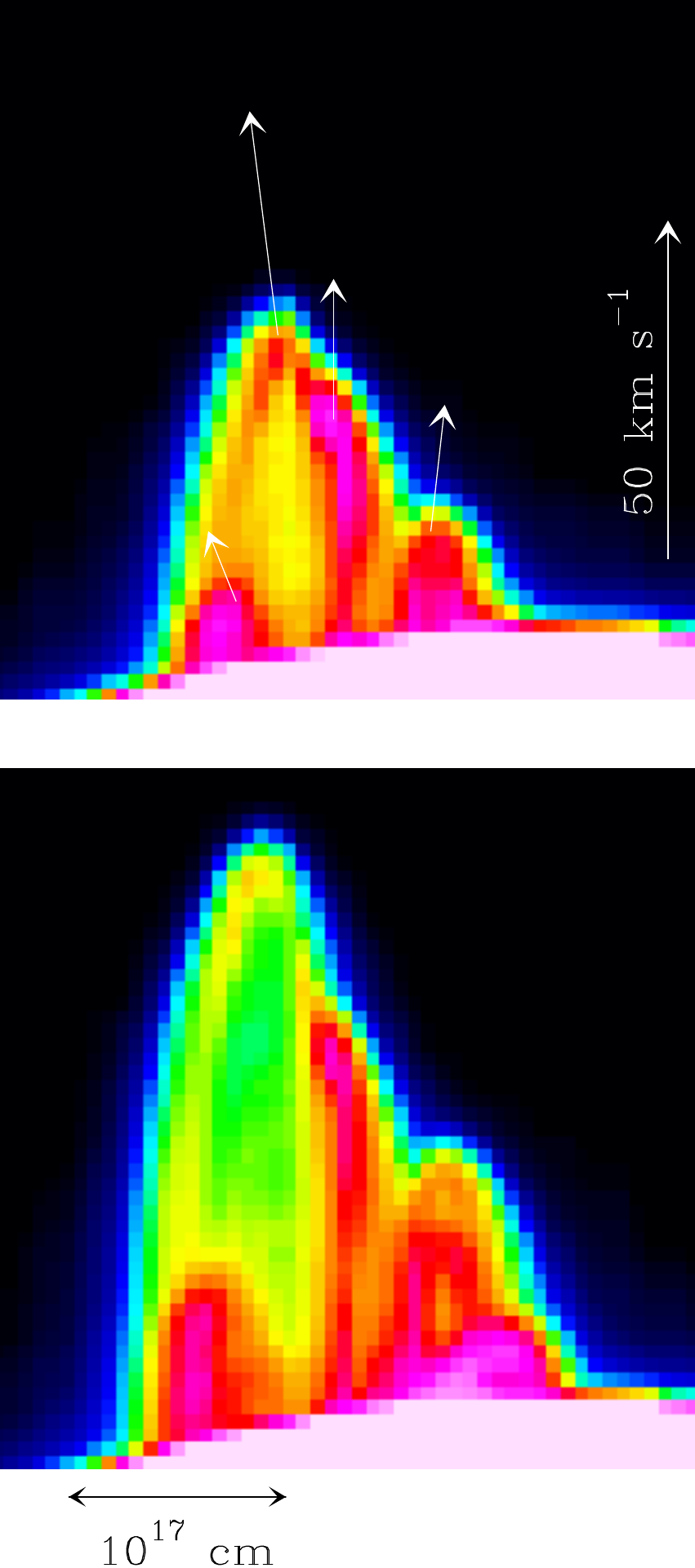}
\caption{Tangential velocities of the H$_2$ gas as projected on the sky and
based on models 56 and 59 (top). The overall morphology resembles
that of the North lobe of Cep E (bottom).}
\label{fig6}
\end{figure}

\section{Summary}

We have derived proper motions based on six IR images of the Cep E
outflow. Two ground based images obtained in 1996 and 2012 in $H_2$
v=1$-$0 at 2.12~\mum, plus another four obtained 
at 4.5~\mum~with the Spitzer's 
IRAC camera over the 2003--2010 time period.

We have defined four cross-correlation boxes that included
the more compact emission structures. Two boxes for
the tips of the two-ridged cavities (boxes 2 and 3, 
respectively, in Figures 3 \& 4) and those for the Northern
and Southern bowshocks (boxes 1 \& 4, respectively).
 The proper motion of the tip of ridges is complex in both 
the rotational and vibrational emission and
and it may reflect time dependent variation in their illumination
or excitation. On the other hand, the proper motions of the main bowshocks
are well defined; they are moving away from the central
source along the symmetry axis with tangential
velocities of $62.6\pm29.5$ and $94.0\pm26.5$~\kms, respectively.
The Southern bowshock is detected also at optical wavelengths
and is known as HH~377, and its proper motion has been measured
(Noriega-Crespo \& Garnavich 2001) rendering a tangential 
velocity of $(107\pm 14)~$\kms, directed approximately along the outflow
axis. This motion is roughly consistent with the proper motion
that we have obtained for the IR emission of this object.
The proper motions based on the 2.12~\mum~emission 
are about a factor two smaller than those in the mid-IR. 
With a time baseline of $\sim$ 16 yr and a angular resolution
of $\sim 1$\arcsec, a priori one does not have any reason to
believe that this difference in magnitude is not real. If this is
the case, then the offset between the vibrational and rotational H$_2$
emission, plus the difference in velocity, suggests a different
physical 'layer' in the outflow where the vibrational H$_2$ 
gas is excited. That not all the molecular tracers originate 
in the same place in  young stellar outflows, including Cep E, 
has been nicely illustrated by a recent
study of water using {\it Herschel} Space Telecope observations 
(Tafalla et al. 2013).  For Cep E, for instance, the H$_2$O
($2_{12}-1_{01}$ 1670 GHz), CO (J = 2-1) and H$_2$ 
emission (from IRAC 3.6~\mum~channel)
at the same angular resolution (13\arcsec), show a very different
spatial distribution along the flow axis (Tafalla et
al. 2013, Figure 4). In this case, the CO emission peaks closer to 
the source, while the H$_2$O and H$_2$ share the same distribution 
farther away from the source.
This means that gas at a temperature of tens of Kelvins (from CO)
resides at a different place that gas at hundreds of Kelvins (from
H$_2$O and H$_2$). A similar process could be taking place in our
case, where the H$_2$ rotational emission, as measured by IRAC at
4.5\mum (i.e. S(11), S(10) \& S(9) lines) is tracing a higher kinetic 
temperature than the vibrational H$_2$ traced by the 2.12\mum~
emission (see e.g. Giannini et al. 2011, Neufeld et al. 2009) and it 
is moving at a higher velocity as well.

As mentioned in the Introduction, the observed emission at 4.5~\mum~from
young stellar outflows is likely produced by three rotational transitions
of H$_2$. In the case of Cep E, in particular, one expects only a small
contribution from the dust continuum emission at these wavelengths
based on what is observed spectroscopically at 5~\mum~in its North lobe
(Noriega-Crespo et al. 2004b).
Thus the infrared proper motions imply that the molecular Hydrogen
gas in Cep E is moving supersonically, just like its 
atomic/ionic counterpart. That the H$_2$ gas can have large tangential
velocities in young stellar outflows was first noticed 
in the HH~1-2 system (Noriega-Crespo et al. 1997b), where velocities
as high as 400~\kms, comparable to those from the optical tracers, 
were measured. Large H$_2$ flow velocities derived from proper motion
measurements are now certainly not uncommon in young stellar outflows
(Chrysostomou et al. 2000; Caratti o Garatti et al. 2009; Zhang et al. 2013),

The large tangential velocities of H$_2$ gas in Cep E 
or other young stellar outflow are somewhat of a puzzle. 
In these outflows, the bulk of the H$_2$ emission 
arises from collisional excitation due to shocks, although turbulence 
and entrainment may also play a role (Reipurth \& Bally 2001; 
Noriega-Crespo 1997a; Raga el al. 2003). 
The shocks exciting the H$_2$ gas cannot be that strong because 
otherwise the molecule dissociates, and this occurs at shock 
velocities of $\sim 45~$\kms~even in the presence of strong magnetic
($\sim$ 50~$\mu$G) fields (Lepp \& Shull 1983; Draine, 
Roberge \& Dalgarno 1983).
There are, however, models of the H$_2$ emission where higher
shock velocities are possible (Le Bourlot et al. 2002). In these
models, if one allows for an initial magnetic field $\ge$ 
100~$\mu$G, i.e. about a factor 5 or 10 larger than 
what is measure at densities of $10^2$--$10^3$~\cc~in interstellar
clouds (Crutcher et al. 2010), then it is possible to reach
shock velocities as high as 70--80~\kms.
That magnetic fields does play a major role in outflows is nicely
capture by the work on the survival of CO and $H_2$ in magnetized
protostellar disk winds by Panoglou et al. (2012). Although this work
concentrates on scales closer to the launching of the flows
(i.e. with a few au), emphasizes the role of chemistry for the
formation and survival of the molecules. At larger scales, however,
``internal shock waves'' (either J or C-shocks) that arise as a result 
of time variability or instabilities of the flow, 
will control which molecules are
destroyed or reformed (Panoglou et al. 2012).

Since the bright condensations or knots that one observes in the collimated 
outflows are ejected from the driving source nearly ``ballistically'' 
(see e.g. Raga 1993), then the survival of the H$_2$ gas needs to be fine 
tuned in terms of the relative velocities of the ejected gas.
 The relatively simple 3D hydrodynamical numerical simulations
to model Cep E presented here (i.e. without including magnetics fields 
or a chemistry network), although limited, do show tangential 
velocities as high as $\sim 50$~\kms, i.e. near the threshold of H$_2$
dissociation if the gas is being collisionally excited.  Higher shock
velocities, like what we have measured in Cep E, require strong
magnetic fields for the H$_2$ molecules to survive 
(Panoglou et al. 2012), or a very efficient reformation mechanism
(Raga, Williams \& Lim 2005)

\acknowledgements
The authors thank the referees and editors for their careful reading
of the manuscript and their valuable  suggestions. In particular
the realization that rotational H$_2$ can trace higher kinetic
temperatures than the vibrational ones.
This research is based in part on observations made with the 
{\it Spitzer Space Telescope} (NASA contract 1407)  and has made use 
of the NASA/IPAC Infrared  Science Archive, both are operated 
by the Jet Propulsion Laboratory, California Institute of Technology, 
under contract with the National Aeronautics and Space Administration (NASA).


\end{document}